\documentclass[prd,preprint,showpacs,groupedaddress]{revtex4-1}
\usepackage{amsmath}
\usepackage{amsfonts}
\usepackage{amssymb}
\usepackage{geometry}
\usepackage{natbib}
\usepackage[english]{babel}
\usepackage{CJK}
\usepackage{graphicx}
\usepackage{fancyhdr}
\usepackage{epstopdf}
\begin{document}
%\bibliographystyle{apsrev4-1}
%\message{}
  \title{The Varying Speed of Light in Eddington-inspired Born-Infield Gravity with Rainbow Metric}

  \author{Ping Li} \author{Jia-Cheng Ding} \author{Qi-Qi Fan} \author{Xian-Ru Hu} \author{Jian-Bo Deng}\email[Jian-Bo Deng:]{dengjb@lzu.edu.cn}

  \affiliation{Institute of Theoretical Physics, LanZhou University,
    Lanzhou 730000, P. R. China}

  \date{\today}
\begin{abstract}
In this paper, we proposed that at each point of spacetime, in addition to the Riemannian metric tensor which describes the geometry of spacetime and the gravitational field, there is an rainbow metric tensor which shows the geometry of spacetime depends on the energy of the test particle. Because Eddington-inspired Born-Infield(EiBI) gravity can be seen as a special bi-metric gravity, we research a varying speed of light theory for a bi-metric gravity corresponding to EiBI gravity. For FRW universe, we find the energy of test particle will increase with rising energy density of universe. We also research the varying speed of light for three different kinds of rainbow functions and find the sign of parameter $\kappa$ would influence the trend of spend of light.

\end{abstract}



  \maketitle

  \section{Introduction}
Since General Relativity was found by Einstein, some phenomenons were perfectly explained. It not only predicted and proved gravitational red shift and light deflection, but also explained the expansion of universe by the Friedmann equation. However, with the deep research about the evolution about universe, some lacks were gradually revealed. Firstly, the horizon problem and flatness problem showed there may be inflation in early universe. Based on the idea, Guth put forward original  inflation mode which take advantage of the thought of grand unified theory, but magnetic monopole problem can not completely solved in the theory. Henceforth, all kind of deformed gravity theories were gradually present, such as, scalar-tensor theory of gravitation~\cite{R1}, $f(R)$ gravity~\cite{R2}, etc. In addition, although General Relativity predict big-bang theory, the singularity of big-bang still resides in some controversy. In order to settle the problem of singularity, based on Eddington theory~\cite{1,R4} and non-linear electrodynamics of Born and Infeld~\cite{41}, Eddington-inspired Born-Ineld gravity was proposed by M\'{a}ximo Ba\~{n}ados and Pedro G.Ferreira~\cite{2}. Different from the metric Born-Infield-Einstein theory~\cite{3} and the purely affine Eddinton theory\cite{1}, EiBI gravity is a Palatini theory of gravity. This theory~\cite{2} supposes the metric \(g_{\mu \nu}\) and the connection \(\Gamma ^{\alpha}_{\beta \gamma}\) are varied independently and can get a no-singularity universe~\cite{39} or bounding universe~\cite{2}.

 Although all kind of inflation modes brought great success, one interesting question is whether inflation is the right solution to the cosmological puzzles, Is inflation really what nature has chosen to do? Rather than changing the matter content of the Universe, a varying speed of light in the early Universe was show in Refs~\cite{27,38}. In this theory, the Universe matter content is the same as in the General relativity and horizon problem and flatness problem can also solve. Furthermore, based on Special Relativity and modified energy-momentum dispersion relation, the rainbow gravity was developed by Jo\~{a}o Magueijo and Lee Smolin~\cite{14,15,16}. In this theory, the geometry of spacetime depends on the energy of the test particle and observers with different energy would see different geometries of spacetime. Hence, a family of energy-dependent metrics named as rainbow metrics will describe the geometry of spacetime, which is different from general gravity theory. The thermodynamics of black hole have been researched~\cite{z2,z3,z4,z5,z6,z7,z1}. Some researches show that a rainbow gravity can also get no-singularity universe~\cite{17,18} or bounding universe~\cite{19,20} in cosmic background. Moreover, no-singularity universe and bounding universe were research in deformed gravity~\cite{R5,R6}. Through modified energy-momentum dispersion relation, the rainbow gravity  might lead to the varying speed of light(VSL) scenario~\cite{27,38}, which depends on the form of rainbow function.

 We might ask whether a new no singularity theory which can solve horizon problem and flatness problem can be got by combining EiBI gravity and Rainbow gravity. As we all know,  EBI gravity has be showed as a bi-metric theory of gravity through some transformations~\cite{13}, similarly, EiBI gravity  can be also described by a bi-metric theory of gravity. The question above transfers as whether a new no singularity theory which can solve horizon problem and flatness problem can be got by combining bi-gravity gravity and Rainbow gravity. Based on the question, we will start with the article.

  In Sec.\uppercase\expandafter{\romannumeral2}, we review the general ideas about  bi-metric gravity and shows Eddington-inspired Born-Infeld gravity may be a bi-metric gravity. Furthermore, rainbow gravity also be reviewed and its picture of varying speed of light was introduced. In Sec.\uppercase\expandafter{\romannumeral3}, through  respectively considering the two metrics in bi-metric gravity as Riemann metric and rainbow metric, we find some contacts between the energy density of FRW universe and the energy of test particle using three diverse rainbow functions in a act of bi-metric gravity corresponding to the action of EiBI gravity. Sec.\uppercase\expandafter{\romannumeral4} gives a summarization.
\section{The background about Eddington-inspired Born-Infeld gravity and rainbow gravity}
\subsection{Eddington-inspired Born-Infeld gravity}
  According to Ref.\cite{13}, the particular ``bigravity" theory with metrics \([g_{\mu \nu},q_{\mu \nu}]\) was studied with the action
   \begin{equation}\label{302}
   S=\frac{1}{16 \pi G}\int {\sqrt{-g} (K-2\Lambda)+\sqrt{-q} (R-2\Upsilon)+\frac{1}{l^{2}} \sqrt{-q}[\pm q^{\alpha \beta} g_{\alpha \beta}]}+S_{M}[g,q,\Psi] ,
   \end{equation}
   which was studied in~\cite{4,5,6,7,8,9,10,11,12,13}. Here \(\Lambda\) and \(\Upsilon\) are cosmological constants for each sector of the theory. \(K\) and \(R\) are the Ricci scalars for metric \(g_{\mu \nu}\) and \(q_{\mu \nu}\). Varying with respect to \(q_{\mu \nu}\), the equation can be algebraically solved for this field
   \begin{equation}\label{301}
   q_{\mu \nu}=\frac{1}{\Upsilon}(R_{\mu \nu} \pm \frac{1}{l^{2}}g_{\mu \nu}) .
   \end{equation}
  Supposed a special condition \(K=0\), we find the Eq.(\ref{301}) continues to be kept. As a consequence, EiBI gravity would be seen as a kind of bigravity. When we choose plus sign in Eqs.(\ref{302}) and (\ref{301}), we can get
   \begin{equation}\label{303}
     S=\frac{1}{16 \pi G}\int {\sqrt{-g} (-2\Lambda)+\sqrt{-q} (2\Upsilon)}+S_{M}[g,\Gamma,\Psi] ,
   \end{equation}
   when \(\frac{1}{\Upsilon }=\kappa\), \(\Upsilon l^{2}=1\) and \(\Lambda=\frac{\lambda}{\kappa}\), the Eq.(\ref{303}) is the same with the action of \(S_{EiBI}\) which was described as~\cite{2,40,R3} :
\begin{equation}\label{201}
  S_{EiBI}[g,\Gamma,\Psi]=\frac{1}{8 \pi G \kappa}\int d^{4}x [\sqrt{|g_{\mu \nu}+\kappa R_{\mu \nu}|}-\lambda \sqrt{g}]+S_{M}[g,\Gamma,\Psi]   ,
 \end{equation}
 where, \(R_{\mu \nu}\) only depends on connection \(\Gamma_{\mu \nu}^{\sigma}\).
 The equations of motion for this theory are in the following. By varying with respect to the metric  \(g_{\mu \nu}\), one can get
 \begin{equation}
    \frac{\sqrt{|g+\kappa R|}}{\sqrt{|g|}} [(g+\kappa R)^{-1}]^{\mu \nu}-\lambda g^{\mu \nu}=-\kappa T^{\mu \nu} .
 \end{equation}
 By introducing an auxiliary metric \(q_{\mu \nu}\) compatible with \(\Gamma_{\mu \nu}^{\lambda}\), the variation with respect to connection \( \Gamma_{\mu \nu}^{\lambda} \)  can be simplified as
 \begin{equation}
   q_{\mu \nu}=g_{\mu \nu}+\kappa R_{\mu \nu} ,
 \end{equation}
 and \(\Gamma^{\mu}_{\alpha \beta}=\frac{1}{2} q^{\mu \sigma} (q_{\sigma \alpha,\beta}+q_{\sigma \beta,\alpha}+q_{\alpha \beta, \sigma})\). Combining the two equations, one derives the equation
 \begin{equation}
    \sqrt{q} q^{\mu \nu}=\lambda g^{\mu \nu}-\kappa T^{\mu \nu}    .
 \end{equation}

Now we focus on cosmology. Assume that the FRW metric corresponding to \(g_{\mu \nu}\) can be shown as
 \begin{equation}
  ds^{2}=-dt^{2}+a(t)^{2}d\vec{x}\cdot d\vec{x} ,
 \end{equation}
  which describes a homogeneous and isotropic flat universe. The metric couples with ideal fluid \( T^{\mu \nu}=(\rho +p) u^{\mu} u^{\nu}+p g^{\mu \nu} \). The components of the other metric \(q_{\mu \nu}\) can also be assumed as \(q_{00}=-U\) and \(q_{ij}=a^{2}V\delta_{ij}\), so we have
 \begin{equation}\label{1}
  ds_{q}^{2}=-Udt^{2}+a(t)^{2} V d\vec{x}\cdot d\vec{x} .
 \end{equation}
The calculation indicates\cite{2}
\begin{equation}
U=\frac{D}{1+\kappa \rho_{T}},~~~~V=\frac{D}{1-\kappa P_{T}} ,
 \end{equation}
 where \(D=\sqrt{(1+\kappa \rho_{T})(1-\kappa P_{T})^{3}}\), \(\rho_{T}=\rho +\Lambda\), and \(P_{T}=p-\Lambda\).
\subsection{Rainbow gravity}
It is generally believed that the geometry of spacetime is fundamentally described by a quantum theory and the Planck energy plays an important role of threshold separating the classical description from the quantum description. There is a choice considering Planck energy~\cite{14}
 \begin{equation}\label{331}
E^{2}\cdot f^{2}(E/E_{P})-p^{2}\cdotp g^{2}(E/E_{P})=m^{2} ,
 \end{equation}
which is a modified energy-momentum dispersion relation. This can be realized by the action of a non-linear map from momentum space to itself, denoted as \(U:P\rightarrow P\), given by~\cite{15}
 \begin{equation}\label{4}
    U\cdot(E,p_{i})=(U_{0},U_{i})=(f(\frac{E}{E_{P}})E,g(\frac{E}{E_{P}})p_{i}) ,
 \end{equation}
 which implies that momentum space has a non-linear norm, given by
\begin{equation}
   \|p\|^{2}=\eta^{ab} U_{a}(p) U_{b}(p) .
 \end{equation}
 This norm is preserved by a non-linear realization of the Lorentz group, given by
 \begin{equation}\label{6}
   \tilde{~{L^{b}_{a}}}=U^{-1}\cdot L^{b}_{a} \cdot U ,
 \end{equation}
 where \(L^{b}_{a}\) are the usual generators.

 An approach developed in Ref.\cite{37} shows that free field theories in flat spacetime have plane wave solutions, even through the 4-momentum satisfies deformed dispersion relations. For this to be possible the contraction between position and momentum must remain linear. This is
 \begin{equation}
   dx^{a}p_{a}=dx^{0}p_{0}+dx^{i}p_{i}.
 \end{equation}
 If momentum transforms non-linearly (from a non-linear action derived from the generators Eq.(\ref{6})), this requires that the \(dx^a\) transformation is energy-dependent, as explained in~\cite{37}. Based on a map \(U\) in (\ref{4}), it is not difficult to show that the spacetime dual leads to~\cite{14}
   \begin{equation}
  ds^{2}=-\frac{1}{f^{2}}dx_{0}^{2}+\frac{1}{g^{2}}d\vec{x}\cdot d\vec{x} .
 \end{equation}
  When the method was extended to curved spacetime, the deformed FRW metric can be shown\cite{14}
  \begin{equation}\label{8}
  ds^{2}=-\frac{1}{f^{2}}dt^{2}+\frac{1}{g^{2}}a(t)^{2}d\vec{x}\cdot d\vec{x} ,
 \end{equation}
here, \(f\) and \(g\) are energy-dependent rainbow functions.

Generally, the VSL scenario\cite{27,38} is an interesting alternative to cosmological inflation. It was found in \cite{28,29,37} that some deformed dispersion relations might lead to a realization of VSL. According to Eq.(\ref{331}), one makes \(f_{3}=\frac{g(E)}{f(E)}\) and can find (here, the mass of photon \(m=0\))
\begin{equation}
  c=\frac{dE}{dp}=\frac{f_{3}}{1-\frac{Ef_{3}^{'}}{f_{3}}} £¬
\end{equation}
where, \(f_{3}^{'}=\frac{df_{3}}{dE}\).  When \(E\rightarrow 0\) or \(f_{3}=1\), one can find the speed of light is a constant which is \(c_{0}\).
  \section{The relationship between Eddington-inspired Born-Infeld gravity and Rainbow gravity}
  Generally, based on Eq.(\ref{303}), the EiBI gravity can be explained as bigravity with some constraint conditions, however, it is still a question what the physical explanations about the two metrics is. Rosen had proposed that at each point of spacetime, in addition to the Riemannian metric tensor which describes the geometry of spacetime and the gravitational field, there is an Euclidean metric tensor which refers to the flat spacetime and describes the inertial force~\cite{30,31}. In massive gravity, except for general Riemannian metric, a reference metric, corresponding to the background metric around which fluctuations take the Fierz-Pauli form, was introduced in order to construct the interaction term\cite{32,33,34}.

 One might wonder whether there is another choice for these two metrics in the bi-metric gravity. Inspired by rainbow gravity, the one in bi-metric gravity may be rainbow metric may be another metric tensor which shows that different energy's observer would see different geometric of spacetime except for Riemannian metric tensor. As we all know,  Riemannian metric tensor corresponding to General relativity relates to constant speed of light, however, rainbow gravity might relate to VSL theory in some rainbow functions. This would mirror that the universe might be constant spend of light without observers' influence, nevertheless, the speed of light might change when considering the influence of observer.

\begin{figure}[htbp]
     \centering\includegraphics[width=8cm]{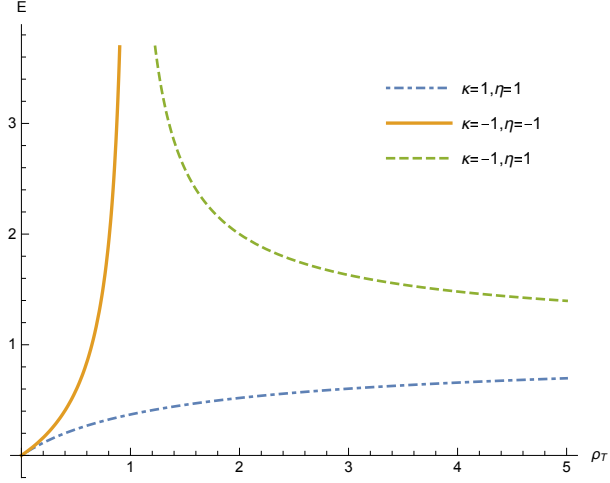}
   \caption{\(E-\rho_{T}\) diagram. It corresponds to \(E_{P}=1\) and \(\eta =1\). From top to bottom, the \(\kappa\) takes respectively \(1\), \(0.1\), \(0.05\).}\label{F1}
\end{figure}

Currently, through some researches of physicists, three rainbow functions were put forward~\cite{25,26}. In order to test the  feasibility for our hypothesis,  the three rainbow functions will be respectively studied. Firstly, originated from loop quantum gravity and noncommutative spacetime\cite{21,22}, the rainbow functions are
\begin{equation}\label{30}
f(E/E_{P})=1,   ~~~~~g(E/E_{P})=\sqrt{1-\eta E/E_{P}} ,
\end{equation}
where \(\eta\) is a model parameter. Based on the hypothesis above, \(U=1\) and \(V=\frac{1}{g(E/E_{P})^{2}}\) can be given, so the relation between the energy of test particle and energy density of universe can be shown as
\begin{equation}\label{31}
E=\frac{E_{P}}{\eta}(1-\frac{1}{(1+\kappa \rho_{T})^{2/3}}) ,
\end{equation}
which is described in Fig.(\ref{F1}). Eq.(15) indicates when \( (\kappa \rho_{T})^{2/3} \) tends to zero, the energy of test particle tends to zero no matter what these parameters $\eta$ and $\kappa$ are. Fig.(1) shows the energy of test particle is monotonously increasing with  $\eta>0$ and $\kappa>0$.  Nevertheless, when $\kappa<0$, there is a critical point $\rho_{T}=-1/\kappa$ which correspond to infinite energy of test particle. The energy of test particle is monotonously increasing with  $\eta>0$ and $\kappa<0$. Moreover, it is monotonously decreasing with  $\eta>0$ and $\kappa>0$, but in following we will find it corresponds to negative speed of light.

Generally, the density parameter \(\Omega\) was showed as
\begin{equation}
    \Omega=8 \pi G \rho/3H^{2} .
\end{equation}
Here, present Hubble parameter \(H_{0}=(73.52\pm1.62) km\cdot s^{-1} \cdot Mpc^{-1}=(2.38\pm0.05)\times 10^{-18} s^{-1}\)~\cite{35}. In general, \(G=6.67\times 10^{-11} m^{3}\cdot kg^{-1}\cdot s^{-2}\) and \(k=1.38\times 10 ^{-23} J\cdot K^{-1}\) and Plank energy \(E_{P}=1.959\times 10^{9} J\). So the critical energy density \(\rho_{C} \approx (1.01\pm0.04) \times 10^{-26} kg/m^{3}\). Suppose that the energy of test particle is proportional to cosmic microwave background radiation (CMBR) temperature \(E=k T\). Because \(\rho_{T}\) includes cosmological constant, so we suppose \(\Omega=\rho_{T}/\rho_{c}=1\) and \(\rho_{T}=\rho_{c}\). Using taylor expansion for Eq.(\ref{31}) when \(E/E_{P}\) or \(\kappa \rho_{T}\) is smaller than 1
\begin{equation}
kT\approx\frac{E_{P}}{\eta} \frac{2}{3}\kappa \rho_{T} .
\end{equation}
The constraint between \(\kappa\) and parameter \(\eta\) can be given
\begin{equation}
\frac{\kappa}{\eta} \approx \frac{4 \pi G kT}{\Omega E_{P} H^{2}}=2.8345 \times 10^{-6} m^{3}/kg  ,
\end{equation}
here, we suppose the energy of test particle is proportional to the temperature of CMBR \(T\cong  2.72548\pm 0.00057 K\)~\cite{36}. Due to \(E\leq E_{P}\), so we can get \(\eta \geq 1\) with \(\kappa \leq 2.8345 \times 10^{-6} m^{3}/kg \). As a result, \(\kappa \rho_{T}\leq 2.8628\times 10^{-32} \) for current universe.

\begin{figure}[htbp]
     \centering\includegraphics[width=8cm]{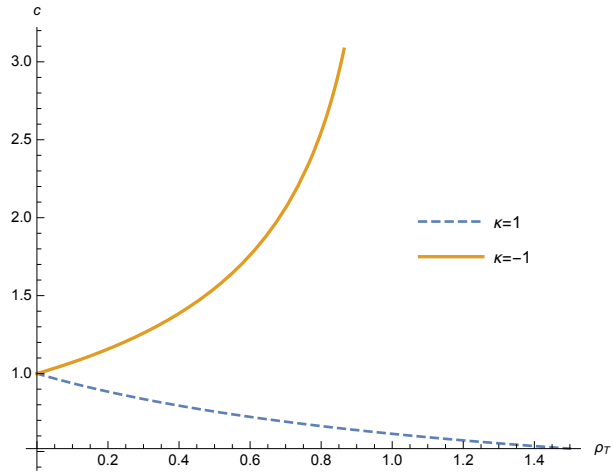}
   \caption{\(c-\rho_{T}\) diagram. It corresponds to \(E_{P}=1\) and \(\kappa =1\).}\label{c1}
\end{figure}

 From rainbow function of Eq.(\ref{30}) we have
\begin{equation}
  c=\frac{2}{(1+\kappa \rho_{T})+(1+\kappa \rho_{T})^{1/3}} ,
\end{equation}
 which showed in Fig.\ref{c1}. In order to get positive speed of light $c$ there must be $\rho_{T}<1$ from Eq.(24). One can find the speed of light is  monotonously decreasing from $1$ to $0$ with $\kappa>0$. It is  monotonously increasing from $1$ to infinity with $\kappa<0$ when $<0\rho_{T}<1$. Moreover, for current universe, we can find \(c\approx (1-1.9085\times 10^{-32}) c_{0}\).
%%%%%%%%%%%%%%%%%%%%%%%%%%%%%%%%%%%%%%%%%%%%%%%%%

\begin{figure}[htbp]
     \centering\includegraphics[width=8cm]{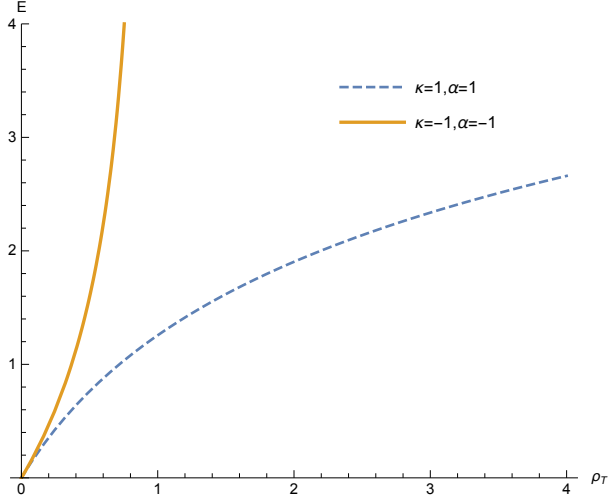}
   \caption{\(E-\rho_{T}\) diagram. It corresponds to \(E_{P}=1\) and \(\kappa =1\). From top to bottom, the \(\alpha\) takes respectively \(2\), \(1\), \(0.5\).}\label{F2}
\end{figure}

Secondly, in order to explain the hard spectra of gamma-ray bursts at cosmological distances, the rainbow functions are proposed to be\cite{23}
\begin{equation}\label{300}
  f(E/E_{P})=\frac{\exp (\alpha E/E_{P})-1}{\alpha E/E_{P}},   ~~~~~g(E/E_{P})=1 ,
 \end{equation}
 where \(\alpha\) is a model parameter. Similar to above discussion, \(U=\frac{1}{f(E/E_{P})^{2}}\) and \(V=1\) can lead to
\begin{equation}\label{32}
\kappa \rho_{T}=\frac{\exp (\alpha E/E_{P})-1}{\alpha E/E_{P}}-1 ,
\end{equation}
 which is described in Fig.(\ref{F2}). Through Fig.\ref{F2}, we can find that \(E\) will also increase monotonically with rising \(\rho_{T}\) with arbitrary $\kappa$, but the changing rate for $\kappa<0$ and $\alpha<0$ is steeper than $\kappa>0$ and $\alpha>0$. When \(\kappa \rho_{T} \) tends to zero, the energy \(E\) tends to zero. In addition, similar to Fig.\ref{F1}, we can find $\kappa\rho_{T}>1$ with $\kappa<0$ and $\kappa\rho_{T}=1$ correspond to diverging energy of test particle.

\begin{figure}[htbp]
     \centering\includegraphics[width=8cm]{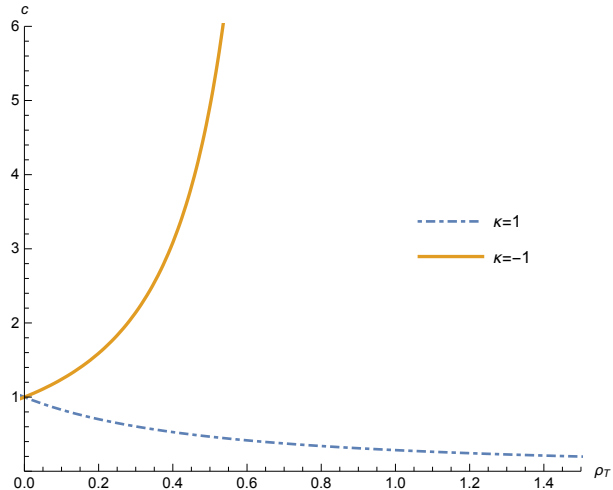}
   \caption{\(E-\rho_{T}\) diagram. It corresponds to \(E_{P}=1\) and \(\kappa =1\). From top to bottom, the \(\alpha\) takes respectively \(2\), \(1\), \(0.5\).}\label{c2}
\end{figure}

Using taylor expansion for Eq.(\ref{32}) when \(E/E_{P}\) or \(\kappa \rho_{T}\) is smaller than 1, we can get
\begin{equation}
 kT\approx\frac{E_{P}}{\alpha} 2\kappa \rho_{T} .
\end{equation}
The constraint between \(\kappa\) and parameter \(\alpha\) can be given
\begin{equation}
\frac{\kappa}{\alpha} \approx \frac{4 \pi G kT}{3 \Omega E_{P} H^{2}}=0.9448 \times 10^{-6} m^{3}/kg .
\end{equation}
Suppose that \(\alpha \geq 1\) ,  we can obtain \(\kappa \leq  0.9448 \times 10^{-6} m^{3}/kg\). As a result,  \(\kappa \rho_{T} \leq 0.9542 \times 10^{-32} \) for present universe.
For the rainbow function of Eq.(\ref{300}), the speed of light can be expressed as
\begin{equation}
  c=\exp{(-\frac{\alpha E}{E_{P}})} .
\end{equation}
From Eq.(26) and Eq.(29), the change of speed of light with the energy density is showed in Fig.\ref{c2}. In order to get positive speed of light $c$ there must be $\rho_{T}<1$ from Eq.(24). One can find the speed of light is  monotonously decreasing from $1$ to $0$ with $\kappa>0$. It is  monotonously increasing from $1$ to infinity with $\kappa<0$ when $0<\rho_{T}<1$. For present universe \(c\approx (1-1.9084 \times 10^{-32}) c_{0}\) and \(c_{0}\approx c \).
%%%%%%%%%%%%%%%%%%%%%%%%%%%%%%%%%%%%%%%%%%%%%%%%%%%%%%%%%%%%

\begin{figure}[htbp]
     \centering\includegraphics[width=8cm]{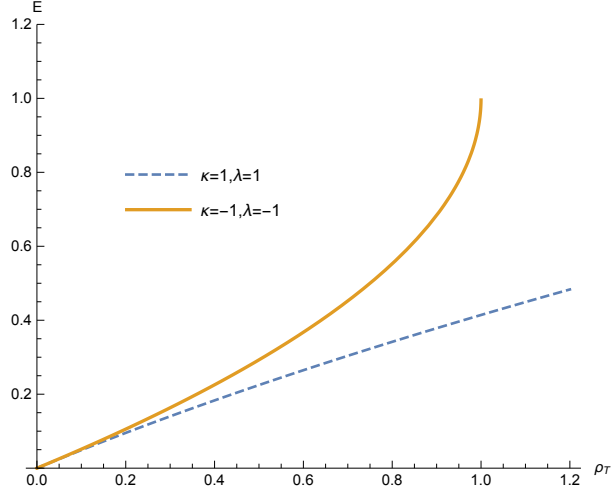}
   \caption{\(E-\rho_{T}\) diagram. It corresponds to \(E_{P}=1\) and \(\lambda =1\). From top to bottom, the \(\kappa\) takes respectively \(1\), \(0.1\), \(0.05\).}\label{F3}
\end{figure}

Thirdly, providing a constant speed of light and a solution to the horizon problem, the rainbow functions are proposed to be~\cite{24}
\begin{equation}\label{330}
  f(E/E_{P})=g(E/E_{P})=\frac{1}{1+\lambda E/E_{P}} ,
\end{equation}
where \(\lambda\) is a model parameter and may have either sign\cite{15}. When \(U=V=\frac{1}{f(E/E_{P})^{2}}=\frac{1}{g(E/E_{P})^{2}}\), we can get
\begin{equation}\label{333}
E=\frac{E_{P}}{\lambda}(\sqrt{1+\kappa \rho_{T}}-1) ,
\end{equation}
which is described in Fig.(\ref{F3}). From Eq.(24), we can find there must be $\kappa\rho_{T}>-1$. Fig.(\ref{F3}) shows that \(E\) will also increase monotonically with rising \(\rho_{T}\) with arbitrary $\kappa$, but the changing rate for $\kappa<0$ and $\alpha<0$ is steeper than $\kappa>0$ and $\alpha>0$. In addition, for $\kappa<0$ and $\alpha<0$, there is a critical point $\kappa\rho_{T}=-1$ which corresponds to a maximum value for the energy of test particle. When \(\kappa \rho_{T} \) tends to zero, the energy \(E\) tends to zero.

Using taylor expansion for Eq.(\ref{333}) when \(E/E_{P}\) or \(\kappa \rho_{T}\) is smaller than 1
\begin{equation}
kT\approx\frac{E_{P}}{\lambda} \frac{1}{2}\kappa \rho_{T} .
\end{equation}
The constraint between \(\kappa\) and \(\lambda\) can be given
\begin{equation}
\frac{\kappa}{\lambda} \approx \frac{16 \pi G kT}{3 \Omega E_{P} H^{2}}=3.7793 \times 10^{-6} m^{3}/kg .
\end{equation}
Suppose that \(\lambda \geq 1\) ,  we can also obtain \(\kappa \leq 3.7793 \times 10^{-6}m^{3}/kg \). As a result,  \(\kappa \rho_{T}\leq 3.8171 \times 10^{-32} \) for present universe. Due to \(f_{3}=1\) in Eq.(\ref{330}), the speed of light keeps constant.
\section{Conclusion And Discussions}
In the paper, we proposed that at each point of spacetime, in addition to the Riemannian metric tensor which describes the geometry of spacetime and the gravitational field, there is an rainbow metric tensor which shows the geometry of spacetime depends on the energy of the test particle. Because Eddington-inspired Born-Infield(EiBI) gravity can be seen as a special bi-metric gravity, we research a varying speed of light theory for a bi-metric gravity corresponding to EiBI gravity. Based on above hypothesis, the three forms of rainbow function have been respectively studied.

 Firstly, based on three rainbow functions, we get three relations between the energy of test particle \(E\) and the energy density of universe  \(\rho_{T}\) and find that the energy of test particle is increasing with rising energy density for all of them. When \(\rho _{T}\rightarrow 0\), \(E\rightarrow0\) and \(\rho _{T}\rightarrow \infty\), \(E\rightarrow \frac{E_{P}}{\eta}\),\(E\rightarrow \frac{E_{P}}{\alpha}\),\(E\rightarrow \frac{E_{P}}{\lambda}\). These results indicate the energy density \(\rho_{T}\) may be sizable in early universe and may be extremely small in current universe.

Secondly, because some features of evolution of universe are reflected by the temperature of CMBR, we suppose that the energy of test particle \(E=k T\) (here, the temperature \(T\) equals to the temperature of CMBR),  which shows three constraints between \(\kappa\) and other parameters (\(\eta\), \(\alpha\) and \(\lambda\)) . In this paper, due to smeller energy density in current universe, we get three approximate expression about them. The restriction would give a way to test the validity of EiBI gravity or bi-metric gravity.

Thirdly, based on three rainbow functions, we research whether the speed of light would be influenced by them. For two conditions of them (Eqs.(\ref{30}) (\ref{300})), we find those speeds of light are increasing with the evolution of universe and only depend on cosmic time when $\kappa>0$. Moreover, those speeds of light are decreasing with the evolution of universe when $\kappa<0$. In early universe, the speed of light is smeller than current state, but in current universe the speed of light approaches to \(c_{0}\). In particular, the rate of change about the speed of light is negligible in current universe. The solutions allow us to imagine that the speed of light is approximate constant in current universe. For another rainbow function Eq.(\ref{330}), the speed of light keeps constant. Based on above researches, we find our assumption might be a right choice to study EiBI gravity.

%%%%%%%%%%%%%%%%%%%%%%%%%%%%%%%%%%%%%%%%%%%%%%%%%%%%%%5
\section*{Conflicts of Interest}
The authors declare that there are no conflicts of interest regarding the publication of this paper.
\section*{Acknowledgments}
 We would like to thank the National Natural Science Foundation of China (Grant No.11571342) for supporting us on this work.

\section*{References}

 \bibliographystyle{unsrt}
  \bibliography{reference}

\end{document}